# Emergence of Self-Awareness in Artificial Systems: A Minimalist Three-Layer Approach to Artificial Consciousness

**Kurando IIDA**

CEO, ErudAite Inc.

Tokyo, Japan

**Abstract**

This paper proposes a minimalist three-layer model for artificial consciousness, focusing on the emergence of self-awareness. The model comprises a Cognitive Integration Layer, a Pattern Prediction Layer, and an Instinctive Response Layer, interacting with Access-Oriented and Pattern-Integrated Memory systems. Unlike brain-replication approaches, we aim to achieve minimal self-awareness through essential elements only. Self-awareness emerges from layer interactions and dynamic self-modeling, without initial explicit self-programming. We detail each component's structure, function, and implementation strategies, addressing technical feasibility. This research offers new perspectives on consciousness emergence in artificial systems, with potential implications for human consciousness understanding and adaptable AI development. We conclude by discussing ethical considerations and future research directions.

**Keywords:** Artificial Consciousness, Self-Awareness, Cognitive Architecture, Emergent Properties

# 1. Introduction

The pursuit of artificial consciousness (AC) represents one of the most ambitious and challenging frontiers in artificial intelligence (AI) research. While significant strides have been made in developing sophisticated AI systems capable of complex pattern recognition and decision-making, the creation of truly conscious artificial entities remains elusive. This paper proposes a novel approach to realizing artificial consciousness, focusing on the emergence of self-awareness through a minimalist three-layer model.

The primary objective of this research is not to faithfully replicate the human brain or consciousness in its entirety. Instead, we aim to achieve minimal 'self-awareness' with minimal constituent elements. This approach is rooted in the belief that consciousness, particularly self-awareness, can emerge from the interaction of simpler cognitive processes and structures.

Our model's uniqueness lies in its potential to allow AI systems to acquire 'self-recognition' without explicitly programming an initial concept of 'self'. This stands in contrast to many existing approaches that either attempt to hard-code self-awareness or rely on extremely complex neural networks to mimic brain function. By focusing on the fundamental processes that may give rise to self-awareness, we aim to provide new insights into both artificial and biological consciousness.

The proposed three-layer model consists of a Cognitive Integration Layer (CIL), a Pattern Prediction Layer (PPL), and an Instinctive Response Layer (IRL). These layers interact with two distinct memory systems: Access-Oriented Memory (AOM) and Pattern-Integrated Memory (PIM). Through the intricate interactions of these components, we hypothesize that a form of self-awareness can emerge.

While our model draws inspiration from neuroscientific understanding of brain function, it does not aim to replicate specific brain structures. Instead, it focuses on implementing key functional aspects that we believe are essential for the emergence of self-awareness. Importantly, our approach ensures that the model remains technically feasible and does not contradict known principles of brain function.

The potential implications of this research extend beyond the field of AI. By providing a new framework for understanding how self-awareness might emerge from simpler processes, this work may offer valuable insights into human consciousness. Furthermore, the structures and processes proposed in this model may correspond to yet-undiscovered aspects of brain function, potentially guiding future neuroscientific research.

In the following sections, we will first provide a comprehensive theoretical background, detailing the concepts of consciousness and self-awareness, existing approaches to artificial consciousness, and the rationale behind our minimalist approach. We will then describe our three-layer model in detail, including the structure and function of each layer and memory system. The emergence of self-awareness within this model will be thoroughly examined, followed by a discussion of technical feasibility and implementation considerations. Finally, we will explore the implications of this research and potential future directions.

By presenting this minimalist approach to artificial consciousness, we hope to stimulate new directions in AI research, contribute to our understanding of consciousness, and pave the way for the development of AI systems with genuine self-awareness.

# 2. Theoretical Background
## 2.1 Concepts of Consciousness and Self-Awareness

The study of consciousness has long been a subject of philosophical inquiry, psychological investigation, and more recently, neuroscientific research. Despite significant advances, a unified definition of consciousness remains elusive. For the purposes of this study, we focus specifically on self-awareness as a key component of consciousness.

Philosophically, self-awareness has been described by thinkers such as John Locke and David Hume as the mind's ability to introspect and examine its own thoughts. In psychology, William James proposed the concept of the "stream of consciousness," emphasizing the continuous flow of thoughts and self-reflection.

From a neuroscientific perspective, self-awareness is often associated with activity in the prefrontal cortex, posterior cingulate cortex, and insula (Kjaer et al., 2002). However, it's important to note that identifying neural correlates does not fully explain the emergence of self-awareness.

In this research, we adopt an operational definition of self-awareness as the system's ability to:

1. Recognize its own existence as distinct from its environment
2. Monitor and reflect on its internal states and processes
3. Understand the consequences of its actions on itself and its environment

This definition allows for a more concrete approach to implementing and evaluating self-awareness in artificial systems.

## 2.2 Existing Approaches to Artificial Consciousness

Efforts to create artificial consciousness have broadly fallen into three categories:

1. **Symbolic AI Approaches:** These attempts, rooted in the tradition of Good Old-Fashioned AI (GOFAI), seek to create consciousness through rule-based systems and symbolic manipulation. Examples include LIDA (Franklin et al., 2014) and ACT-R (Anderson et al., 2004). While these models offer explicit representations of cognitive processes, they often struggle with the flexibility and adaptability observed in human consciousness.

2. **Neural Network-Based Approaches:** With the advent of deep learning, researchers have explored the possibility of consciousness emerging from complex neural networks. Models like Cleeremans' Radical Plasticity Thesis (2011) suggest that consciousness arises from the brain's ability to learn to be conscious. However, these approaches often lack interpretability and struggle to explain specific aspects of conscious experience.

3. **Hybrid Approaches:** Recognizing the limitations of pure symbolic or connectionist models, some researchers have proposed hybrid systems. The Global Workspace Theory (Baars, 1997) implemented in computational models (e.g., Shanahan, 2006) is an example of this approach. These models attempt to combine the strengths of both symbolic and neural network paradigms but often result in increased complexity.

Each of these approaches has contributed valuable insights to the field of artificial consciousness. However, they often either rely on overly complex systems or fail to address the emergence of self-awareness as a fundamental aspect of consciousness.

## 2.3 Minimalist Approaches in AI and Cognitive Science

The concept of minimalism in cognitive science and AI research stems from the principle of Occam's razor—the idea that the simplest explanation is often the correct one. In the context of consciousness studies, minimalist approaches seek to identify the most fundamental components necessary for the emergence of conscious-like behaviors.

The Minimal Model of Consciousness, proposed by Morsella et al. (2016), suggests that consciousness arises from the conflict between different action plans. This model emphasizes the role of consciousness in action selection, providing a streamlined framework for understanding basic conscious processes.

In AI, minimalist approaches have been successful in various domains. For instance, the success of simple reinforcement learning algorithms in complex tasks (e.g., AlphaGo by Silver et al., 2017) demonstrates that sophisticated behaviors can emerge from relatively simple principles.

The advantages of minimalist approaches include:
1. Improved interpretability and transparency
2. Reduced computational complexity
3. Easier identification of core principles and mechanisms
4. Greater potential for generalization across different domains

However, minimalist approaches also face challenges, particularly in capturing the full richness and complexity of human-like consciousness. The key is to strike a balance between simplicity and explanatory power.

Our proposed three-layer model builds upon these minimalist principles, aiming to create a system capable of self-awareness using the simplest possible architecture that can account for this phenomenon. By focusing on the essential components and their interactions, we hope to shed light on the fundamental processes underlying the emergence of self-awareness, both in artificial systems and potentially in biological entities.

In the following sections, we will detail how our model implements these minimalist principles while addressing the limitations of existing approaches to artificial consciousness.

# 3. The Three-Layer Model: A Minimalist Approach
## 3.1 Overview of the Model
Our proposed model for artificial consciousness consists of three primary layers: the Cognitive Integration Layer (CIL), the Pattern Prediction Layer (PPL), and the Instinctive Response Layer (IRL). These layers interact with two memory systems: Access-Oriented Memory (AOM) and Pattern-Integrated Memory (PIM). The model is designed to be minimalist, focusing on the essential components necessary for the emergence of self-awareness.

The key principles guiding the design of this model are:
1. **Modularity:** Each layer has a specific function, allowing for clear analysis of its role in the emergence of self-awareness.
2. **Interactivity:** The layers and memory systems interact dynamically, creating complex behaviors from simple components.
3. **Adaptability:** The model can learn and adapt to new information and environments.
4. **Scalability:** The basic structure can be expanded or refined without fundamentally altering its core principles.

The overall structure of the model can be visualized as follows:

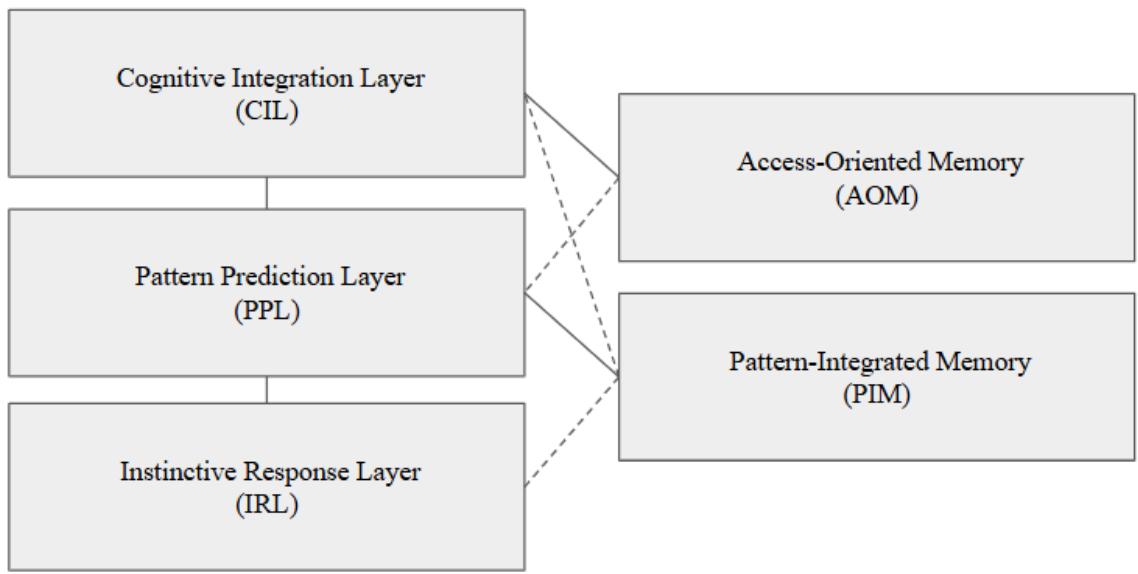

Each layer and memory system plays a crucial role in the emergence of self-awareness:
1. **CIL:** Responsible for integrating information from other layers, making high-level decisions, and managing the system's overall cognitive state.
2. **PPL:** Focuses on recognizing patterns and making predictions based on input data and past experiences.

3. **IRL:** Handles basic, instinctive responses to stimuli and maintains the system's core operational state.
4. **AOM:** Provides rapid access to specific information and experiences.
5. **PIM:** Stores and integrates patterns and abstract knowledge learned over time.

In the following sections, we will delve into the details of each component, explaining their structure, function, and implementation methods.

## 3.2 Cognitive Integration Layer (CIL)

The Cognitive Integration Layer (CIL) serves as the central executive of our model, analogous to the role of the prefrontal cortex in the human brain. Its primary functions include:

1. Information integration from other layers
2. High-level decision making
3. Self-monitoring and introspection
4. Management of goal-directed behavior

**Structure and Function**

The CIL is structured as a graph-based network where nodes represent concepts, ideas, or states, and edges represent relationships between them. This structure allows for flexible representation of complex cognitive states and relationships.

Key components of the CIL include:

- **Concept Graph:** A dynamic graph structure representing the system's current understanding and cognitive state.
- **Attention Mechanism:** A system for focusing on relevant information and ignoring distractions.
- **Goal Management System:** A component for setting, prioritizing, and pursuing goals.
- **Self-Model:** A specialized subgraph representing the system's model of itself.

**Implementation Method**

The implementation of the CIL involves several steps:

1. **Graph Database Selection:**
   - Choose a graph database system (e.g., Neo4j, ArangoDB) for storing the concept graph.
   - Implement a Python interface for graph operations using libraries like *py2neo* or *pyArango*.
2. **Node and Edge Definition:**
   - Define node types (e.g., Concept, State, Goal) with attributes such as activation level, creation time, and last access time.
   - Define edge types (e.g., IsA, HasProperty, Causes) with attributes like strength and

directionality.

3. **Labeling Algorithm:**
   - Implement a natural language processing pipeline using libraries like *spaCy* or *NLTK* for text input processing.
   - Develop a custom labeling function that:
     
     a. Extracts key concepts from input data
     
     b. Compares new concepts with existing nodes using cosine similarity of word embeddings
     
     c. Creates new nodes or updates existing ones based on similarity thresholds

   **Example pseudo-code for the labeling function:**

   ```
   def label_input(input_text, concept_graph):
       concepts = extract_concepts(input_text)
       for concept in concepts:
           embedding = get_word_embedding(concept)
           similar_nodes = find_similar_nodes(concept_graph, embedding)
           if max_similarity(similar_nodes) > SIMILARITY_THRESHOLD:
               update_existing_node(concept_graph, similar_nodes[0], concept)
           else:
               create_new_node(concept_graph, concept, embedding)
   ```

4. **Similarity Calculation:**
   - Implement cosine similarity function for comparing word embeddings.
   - Use a pre-trained word embedding model (e.g., Word2Vec, GloVe) for initial concept representations.

5. **Graph Exploration Algorithms:**
   - Implement depth-first and breadth-first search algorithms for exploring the concept graph.
   - Develop a spreading activation algorithm for concept retrieval and association.

6. **Attention Mechanism:**
   - Implement a priority queue data structure for managing attentional focus.
   - Develop an attention scoring function based on factors like concept activation, relevance to current goals, and novelty.

7. **Goal Management System:**
   - Create a goal class with attributes such as priority, deadline, and subgoals.
   - Implement a goal selection algorithm that considers current system state and environmental factors.

8. **Self-Model:**
    - Create a specialized subgraph structure for representing the system's self-model.
    - Develop update mechanisms that modify the self-model based on system experiences and feedback.

By implementing these components, the CIL can effectively integrate information, make decisions, and maintain a dynamic model of itself and its environment. This layer plays a crucial role in the emergence of self-awareness by providing a structure for self-representation and introspection.

In the next section, we will discuss the Pattern Prediction Layer (PPL) and its implementation.

## 3.3 Pattern Prediction Layer (PPL)

The Pattern Prediction Layer (PPL) is designed to recognize patterns in input data and make predictions about future states or events. This layer is crucial for the system's ability to learn from experience and anticipate outcomes, which are key components of intelligent behavior and self-awareness.

**Structure and Function**

The PPL is structured as a deep neural network, specifically utilizing a Transformer architecture due to its effectiveness in capturing long-range dependencies and its ability to handle various types of input data. The main functions of the PPL include:

1. Pattern recognition in multi-modal input data
2. Sequence prediction and generation
3. Feature extraction and representation learning
4. Anomaly detection and novelty recognition

**Implementation Method**

The implementation of the PPL involves the following steps:

1. **Architecture Selection:**
    - Implement a Transformer-based architecture using a deep learning framework such as PyTorch or TensorFlow.
    - Design the model to handle multi-modal inputs (e.g., text, numerical data, categorical data).
2. **Input Processing:**
    - Develop preprocessing pipelines for each input modality.
    - Implement tokenization for text data, normalization for numerical data, and encoding for categorical data.
3. **Model Structure:**
    - Implement multi-head self-attention mechanisms.
    - Design feed-forward neural networks for each Transformer block.

o   Implement positional encoding to maintain sequence order information.

**Here's a simplified pseudo-code for the Transformer block:**

```
class TransformerBlock(nn.Module):
    def __init__(self, d_model, n_heads, d_ff, dropout):
        super().__init__()
        self.attention = MultiHeadAttention(d_model, n_heads)
        self.norm1 = LayerNorm(d_model)
        self.ff = FeedForward(d_model, d_ff)
        self.norm2 = LayerNorm(d_model)
        self.dropout = nn.Dropout(dropout)

    def forward(self, x):
        attended = self.attention(x)
        x = self.norm1(x + self.dropout(attended))
        fedforward = self.ff(x)
        return self.norm2(x + self.dropout(fedforward))
```

4. **Training Algorithm:**
    o   Implement a self-supervised learning approach, such as masked language modeling for text data or contrastive predictive coding for multi-modal data.
    o   Use adaptive learning rate algorithms like Adam optimizer.
    o   Implement gradient clipping to prevent exploding gradients.

5. **Prediction Mechanism:**
    o   Develop a beam search algorithm for generating multiple likely continuations of a given input sequence.
    o   Implement temperature sampling for controlling the randomness of predictions.

**Example pseudo-code for beam search:**

```
def beam_search(model, initial_sequence, beam_width, max_length):
    sequences = [(initial_sequence, 0)]
    for _ in range(max_length):
        candidates = []
        for seq, score in sequences:
            predictions = model.predict(seq)
            top_k = get_top_k(predictions, k=beam_width)
            for token, prob in top_k:
                new_seq = seq + [token]
                new_score = score + log(prob)
                candidates.append((new_seq, new_score))
        sequences = sorted(candidates, key=lambda x: x[1], reverse=True)[:beam_width]
    return sequences[0][0]
```

6. **Feature Extraction:**
    - Utilize the attention mechanisms and intermediate layer outputs as feature extractors.
    - Implement dimensionality reduction techniques (e.g., PCA, t-SNE) for visualizing learned representations.

7. **Anomaly Detection:**
    - Develop a mechanism to calculate the surprise or perplexity of input sequences.
    - Implement thresholding to identify anomalous or novel inputs.

8. **Model Optimization:**
    - Implement techniques like knowledge distillation to create smaller, more efficient versions of the model.
    - Use quantization and pruning to reduce model size and increase inference speed.

By implementing these components, the PPL can effectively recognize patterns, make predictions, and extract meaningful features from input data. This layer contributes to the emergence of self-awareness by providing the system with the ability to anticipate and understand its environment, as well as recognize novel or unexpected situations.

### 3.4 Instinctive Response Layer (IRL)

The Instinctive Response Layer (IRL) is designed to handle basic, rapid responses to stimuli and maintain the system's core operational state. This layer is analogous to the limbic system and brainstem in biological organisms, providing quick, reflexive behaviors and regulating fundamental system states.

**Structure and Function**

The IRL is structured as a combination of rule-based systems and reinforcement learning models. Its main functions include:

1. Rapid response to urgent stimuli
2. Maintenance of system homeostasis
3. Generation of basic emotional states
4. Prioritization of survival-related behaviors

**Implementation Method**

The implementation of the IRL involves the following steps:

1. **State Space Definition:**
    - Define a set of core system states (e.g., energy level, integrity level, operational mode).
    - Implement a state monitoring system that continuously updates these values.
2. **Action Space Definition:**
    - Define a set of basic actions the system can take (e.g., rest, seek resource, avoid threat).
    - Implement action execution functions for each defined action.
3. **Rule-Based System:**
    - Develop a set of if-then rules for immediate responses to critical situations.
    - Implement a rule execution engine that can quickly evaluate and trigger appropriate responses.

**Example pseudo-code for the rule-based system:**

```
class Rule:
    def __init__(self, condition, action):
        self.condition = condition
        self.action = action

class RuleEngine:
    def __init__(self, rules):
        self.rules = rules

    def evaluate(self, state):
        for rule in self.rules:
            if rule.condition(state):
                rule.action(state)
```

4. **Reinforcement Learning Model:**
    - Implement a Q-learning or SARSA algorithm for learning optimal behaviors in various states.
    - Define a reward function that encourages system survival and stability.

**Here's a simplified implementation of Q-learning:**

```python
class QLearning:
    def __init__(self, states, actions, learning_rate, discount_factor):
        self.q_table = {s: {a: 0 for a in actions} for s in states}
        self.lr = learning_rate
        self.gamma = discount_factor

    def update(self, state, action, reward, next_state):
        current_q = self.q_table[state][action]
        max_next_q = max(self.q_table[next_state].values())
        new_q = current_q + self.lr * (reward + self.gamma * max_next_q - current_q)
        self.q_table[state][action] = new_q

    def get_best_action(self, state):
        return max(self.q_table[state], key=self.q_table[state].get)
```

5. **Emotional State Generator:**
   - Implement a basic emotion model (e.g., PAD: Pleasure, Arousal, Dominance).
   - Develop functions to update emotional state based on system state and external stimuli.
6. **Homeostasis Mechanism:**
   - Implement feedback loops for maintaining key system parameters within optimal ranges.
   - Develop adaptive setpoints that can change based on environmental conditions and system experience.
7. **Action Selection Mechanism:**
   - Implement a hybrid action selection system that combines rule-based responses and learned behaviors.
   - Develop a priority system that can override learned behaviors with instinctive responses in critical situations.

**Example pseudo-code for action selection:**

```
def select_action(state, rule_engine, q_learner):
    # Check for critical situations first
    critical_action = rule_engine.evaluate(state)
    if critical_action:
        return critical_action

    # If no critical action, use learned behavior
    return q_learner.get_best_action(state)
```

8. **Integration with Other Layers:**
   - Implement communication channels to send state information and receive modulation signals from other layers.
   - Develop mechanisms for the IRL to influence the attention and decision-making processes in higher layers.

By implementing these components, the IRL can provide rapid, instinctive responses to stimuli while also adapting its behavior through learning. This layer contributes to the emergence of self-awareness by maintaining a basic sense of self-preservation and providing a foundation for more complex cognitive processes in the higher layers.

In the next section, we will discuss how these three layers interact and how their combined operation leads to the emergence of self-awareness.

## 3.5 Interactions Between Layers

The emergence of self-awareness in our model is not the result of any single layer's operation, but rather arises from the complex interactions between the CIL, PPL, and IRL, as well as their interactions with the memory systems (AOM and PIM). These interactions create a dynamic, self-regulating system capable of introspection and self-modeling.

**Key Interaction Pathways**

1. **CIL ↔ PPL:**
   - The CIL sends high-level goals and context to the PPL.
   - The PPL provides predictions and pattern recognition results to the CIL.
2. **CIL ↔ IRL:**
   - The CIL modulates IRL responses based on higher-level goals and context.
   - The IRL sends instinctive responses and emotional states to the CIL.

3. **PPL ↔ IRL:**
    - The PPL provides anticipated outcomes to the IRL.
    - The IRL informs the PPL about basic system states and instinctive reactions.
4. **All Layers ↔ AOM/PIM:**
    - Each layer both reads from and writes to the memory systems.

**Implementation of Layer Interactions**

To implement these interactions, we need to establish robust communication channels between the layers. Here's a detailed approach:

1. **Message Passing System:**
    - Implement a publish-subscribe (pub/sub) messaging system using a library like ZeroMQ or Apache Kafka.
    - Define message types for different kinds of inter-layer communications.

**Example pseudo-code for a basic pub/sub system:**

```
class MessageBus:
    def __init__(self):
        self.subscribers = defaultdict(list)

    def publish(self, topic, message):
        for subscriber in self.subscribers[topic]:
            subscriber(message)

    def subscribe(self, topic, callback):
        self.subscribers[topic].append(callback)

message_bus = MessageBus()

# In CIL
message_bus.subscribe("PPL_predictions", cil.handle_predictions)

# In PPL
message_bus.publish("PPL_predictions", prediction_results)
```

2. **State Synchronization:**
   - Implement a shared state object that all layers can access and modify.
   - Use locks or other concurrency control mechanisms to prevent race conditions.
3. **Callback Mechanisms:**
   - Implement callback functions in each layer to handle incoming messages from other layers.
   - Use event-driven programming to respond to changes in the system state.
4. **Integration Pipeline:**
   - Develop a main integration loop that coordinates the operation of all layers.
   - Implement adaptive time-stepping to balance responsiveness and computational efficiency.

**Here's a simplified example of the main integration loop:**

```
def main_loop(cil, ppl, irl, aom, pim):
    while True:
        # Update system state
        current_state = get_current_state()

        # IRL processing
        instinctive_response = irl.process(current_state)
        message_bus.publish("IRL_response", instinctive_response)

        # PPL processing
        ppl_prediction = ppl.predict(current_state)
        message_bus.publish("PPL_prediction", ppl_prediction)

        # CIL processing
        cil_decision = cil.integrate(current_state, instinctive_response, ppl_prediction)
        message_bus.publish("CIL_decision", cil_decision)

        # Update memories
        aom.update(current_state, cil_decision)
        pim.update(current_state, ppl_prediction)

        # Execute decision
        execute_action(cil_decision)

        # Adjust time step
        adjust_time_step()
```

5. **Conflict Resolution:**
    - Implement a conflict resolution mechanism for cases where different layers suggest conflicting actions.
    - Use a priority system or a voting mechanism to resolve conflicts.
6. **Feedback Loops:**
    - Implement feedback mechanisms that allow each layer to evaluate the outcomes of its decisions and adjust its behavior accordingly.

7. **Meta-cognitive Processes:**
    - Develop mechanisms in the CIL to monitor and evaluate the performance of all layers.
    - Implement adaptive learning rates and parameter tuning based on this meta-cognitive evaluation.

Through these interactions, the system can integrate instinctive responses, pattern-based predictions, and high-level cognitive processes. This integration is key to the emergence of self-awareness, as it allows the system to model its own behavior, predict its own responses, and reflect on its own decision-making processes.

# 4. Memory Systems

The two memory systems, Access-Oriented Memory (AOM) and Pattern-Integrated Memory (PIM), play crucial roles in supporting the operations of the three main layers and in the emergence of self-awareness. These systems provide the substrate for storing experiences, learned patterns, and the evolving self-model.

## 4.1 Access-Oriented Memory (AOM)

The AOM is designed for rapid storage and retrieval of specific information and experiences. It is analogous to episodic memory in biological systems.

**Structure and Function**

The AOM is implemented as a combination of a key-value store and a graph database. This hybrid structure allows for both quick lookups of specific information and the representation of complex relationships between memory elements.

Key components of the AOM include:

1. **Key-Value Store:** For rapid retrieval of individual memory elements.
2. **Graph Structure:** For representing relationships between memories.
3. **Indexing System:** For efficient search and retrieval.
4. **Decay Mechanism:** For managing memory lifespan and relevance.

**Implementation Method**

1. **Data Structure:**
    - Use a distributed key-value store (e.g., Redis) for the primary storage.
    - Implement a graph database (e.g., Neo4j) for storing relationships.
2. **Memory Encoding:**
    - Develop functions to encode various types of information (sensory data, actions, outcomes) into storable formats.
    - Implement a unique identifier generation system for each memory element.

**Example of memory encoding:**

```
def encode_memory(data, type):
    memory_id = generate_unique_id()
    encoded_data = serialize(data)
    timestamp = current_time()
    return {
        'id': memory_id,
        'type': type,
        'data': encoded_data,
        'timestamp': timestamp,
        'access_count': 0
    }
```

3. **Indexing and Search:**
    - Implement an inverted index for keyword-based searches.
    - Use locality-sensitive hashing (LSH) for similarity-based searches.
4. **Memory Retrieval:**
    - Develop a multi-stage retrieval process that combines key-value lookups and graph traversals.
    - Implement a relevance scoring system to rank retrieved memories.

**Example of memory retrieval:**

```
def retrieve_memory(query, context):
    candidate_memories = key_value_store.search(query)
    graph_related_memories = graph_db.find_related(candidate_memories)
    scored_memories = score_relevance(candidate_memories + graph_related_memories, context)
    return sorted(scored_memories, key=lambda x: x['relevance'], reverse=True)
```

5. **Decay Mechanism:**
    - Implement a time-based decay function that reduces the retrieval probability of old memories.
    - Develop a reinforcement mechanism that strengthens frequently accessed memories.

6. **Consolidation Process:**
    - Implement a background process that periodically reviews and reorganizes memories.
    - Develop algorithms for abstracting common patterns from multiple memories and storing them in the PIM.

By implementing these components, the AOM can provide fast, context-sensitive access to specific memories and experiences. This capability is crucial for the system's ability to learn from past experiences, make informed decisions, and develop a sense of continuity that contributes to self-awareness.

## 4.2 Pattern-Integrated Memory (PIM)

The Pattern-Integrated Memory (PIM) is designed to store and integrate patterns, skills, and abstract knowledge learned over time. It is analogous to semantic and procedural memory in biological systems.

**Structure and Function**

The PIM is implemented as a distributed representation system using deep neural networks. Its main functions include:

1. Storage of abstract patterns and concepts
2. Integration of new information with existing knowledge
3. Generation of novel combinations and abstractions
4. Support for generalization and transfer learning

**Implementation Method**

1. **Architecture Selection:**
    - Implement a variational autoencoder (VAE) as the core of the PIM.
    - Use additional layers for task-specific processing (e.g., classification, generation).
2. **Data Representation:**
    - Develop encoding functions to transform various types of input (sensory data, abstract concepts, skills) into vector representations.
    - Implement a hierarchical structure to represent different levels of abstraction.

**Example of a basic VAE structure:**

```python
class VAE(nn.Module):
    def __init__(self, input_dim, latent_dim):
        super(VAE, self).__init__()
        self.encoder = nn.Sequential(
            nn.Linear(input_dim, 256),
            nn.ReLU(),
            nn.Linear(256, 128),
            nn.ReLU()
        )
        self.fc_mu = nn.Linear(128, latent_dim)
        self.fc_var = nn.Linear(128, latent_dim)
        self.decoder = nn.Sequential(
            nn.Linear(latent_dim, 128),
            nn.ReLU(),
            nn.Linear(128, 256),
            nn.ReLU(),
            nn.Linear(256, input_dim)
        )

    def encode(self, x):
        h = self.encoder(x)
        return self.fc_mu(h), self.fc_var(h)

    def reparameterize(self, mu, log_var):
        std = torch.exp(0.5 * log_var)
        eps = torch.randn_like(std)
        return mu + eps * std

    def decode(self, z):
        return self.decoder(z)

    def forward(self, x):
        mu, log_var = self.encode(x)
        z = self.reparameterize(mu, log_var)
        return self.decode(z), mu, log_var
```

3. **Training Process:**
   - Implement a continuous learning process that updates the PIM with new information from the AOM and other layers.
   - Use contrastive learning techniques to improve the quality of learned representations.
4. **Pattern Integration:**
   - Develop algorithms for merging new patterns with existing ones in the latent space.
   - Implement a novelty detection mechanism to identify when new information doesn't fit existing patterns.

**Example of pattern integration:**

```python
def integrate_pattern(pim, new_pattern, learning_rate):
    # Encode the new pattern
    new_encoding = pim.encode(new_pattern)

    # Find similar existing patterns
    similar_patterns = find_similar_patterns(pim, new_encoding)

    # If similar patterns exist, update them
    if similar_patterns:
        for pattern in similar_patterns:
            updated_encoding = interpolate(pattern, new_encoding, learning_rate)
            pim.update_pattern(pattern, updated_encoding)
    else:
        # If no similar patterns, add as a new pattern
        pim.add_new_pattern(new_encoding)
```

5. **Generative Capabilities:**
   - Implement sampling methods to generate new patterns from the learned latent space.
   - Develop interpolation techniques for creating novel combinations of existing patterns.
6. **Hierarchical Structure:**
   - Implement a hierarchical VAE structure to capture different levels of abstraction.
   - Develop mechanisms for information flow between different levels of the hierarchy.
7. **Attention Mechanism:**
   - Implement an attention mechanism to focus on relevant parts of the latent space for

specific tasks.
   - Use this mechanism to guide the generation and retrieval of patterns.
8. **Continual Learning:**
   - Implement techniques like elastic weight consolidation (EWC) to prevent catastrophic forgetting when learning new information.
   - Develop a replay mechanism that revisits and reinforces important past experiences.

**Example of EWC implementation:**

```
class EWC(object):
    def __init__(self, model: nn.Module, dataset: list):
        self.model = model
        self.dataset = dataset
        self.params = {n: p for n, p in self.model.named_parameters() if
p.requires_grad}
        self._means = {}
        self._precision_matrices = self._diag_fisher()

    def _diag_fisher(self):
        precision_matrices = {}
        for n, p in self.params.items():
            p.grad.data.zero_()
            precision_matrices[n] = torch.zeros(p.size())

        self.model.eval()
        for input in self.dataset:
            self.model.zero_grad()
            output = self.model(input).view(1, -1)
            label = output.max(1)[1].view(-1)
            loss = F.nll_loss(F.log_softmax(output, dim=1), label)
            loss.backward()

            for n, p in self.params.items():
                precision_matrices[n].data += p.grad.data ** 2 /
len(self.dataset)

        precision_matrices = {n: p for n, p in precision_matrices.items()}
        return precision_matrices

    def penalty(self, model: nn.Module):
        loss = 0
        for n, p in model.named_parameters():
            _loss = self._precision_matrices[n] * (p - self._means[n]) ** 2
```

By implementing these components, the PIM can effectively store, integrate, and generate patterns and abstract knowledge. This capability is crucial for the system's ability to generalize from past experiences, recognize complex patterns, and develop a rich internal model of the world and itself.

The interaction between the AOM and PIM, facilitated by the three main layers (CIL, PPL, and IRL), creates a dynamic memory system that supports the emergence of self-awareness. The AOM provides specific, episodic-like memories that the system can reflect upon, while the PIM offers a structured representation of the system's accumulated knowledge and skills. Together, they form the basis for the system's evolving self-model and its ability to understand and predict its own behavior in various contexts.

In the next section, we will discuss how these components come together to facilitate the emergence of self-awareness in our artificial consciousness model.

# 5. Emergence of Self-Awareness

The emergence of self-awareness in our model is a result of the complex interactions between the three layers (CIL, PPL, IRL) and the two memory systems (AOM, PIM). This section will detail how these components work together to create a system capable of self-recognition, self-modeling, and self-reflection.

## 5.1 The Role of Labeling in Self-Recognition

Labeling plays a crucial role in the emergence of self-awareness in our model. The process of labeling allows the system to categorize and differentiate between various entities, including itself.

**Implementation of Self-Labeling**

1. **Initial Self-Representation:**
    - Create a special "self" node in the CIL's concept graph.
    - Initialize this node with basic system properties and states.

    ```
    1.    def initialize_self_node(cil):
    2.        self_node = Node("self", type="entity")
    3.        self_node.add_property("system_id", generate_unique_id())
    4.        self_node.add_property("creation_time", current_time())
    5.        cil.concept_graph.add_node(self_node)
    6.        return self_node
    ```

2. **Self-Action Association:**
    - Develop a mechanism to associate actions taken by the system with the "self" node.
    - Implement a feedback loop that updates the self-model based on action outcomes.

    ```
    1.    def associate_action_with_self(cil, action, outcome):
    2.        self_node = cil.get_self_node()
    3.        action_node = Node(action, type="action")
    4.        outcome_node = Node(outcome, type="outcome")
    5.
    6.        cil.concept_graph.add_edge(self_node, action_node,
    ```

3.

4. **Environmental Differentiation:**
    - Implement algorithms to distinguish between changes caused by the system's actions and those caused by external factors.
    - Use this differentiation to reinforce the boundary between "self" and "non-self".

```
1.     def differentiate_self_and_environment(cil, state_change):
2.         if is_result_of_self_action(state_change):
3.             associate_with_self(state_change)
4.         else:
5.             associate_with_environment(state_change)
```

## 5.2 Formation of Self-Concept through Layer Interactions

The formation of a self-concept emerges from the interactions between the three main layers:

1. **CIL Contribution:**
    - Maintains and updates the explicit self-model in the concept graph.
    - Integrates information from other layers to form a coherent self-concept.
2. **PPL Contribution:**
    - Predicts the outcomes of the system's actions, contributing to the sense of agency.
    - Recognizes patterns in the system's behavior, helping to form a consistent self-image.
3. **IRL Contribution:**
    - Provides a basic sense of "self" through instinctive responses and internal state monitoring.
    - Contributes to the emotional aspects of self-awareness.

**Implementation of Self-Concept Formation**

1. **Cross-Layer Information Integration:**
    - Implement a central integration function in the CIL that combines inputs from all layers.

    ```
    1.      def integrate_self_information(cil, ppl_prediction, irl_state):
    2.          self_node = cil.get_self_node()
    3.
    4.          # Update self-node with PPL prediction
    5.          self_node.add_property("predicted_state", ppl_prediction)
    6.
    7.          # Update self-node with IRL state
    8.          self_node.add_property("emotional_state", irl_state.emotion)
    9.          self_node.add_property("physiological_state", irl_state.physiology)
    10.
    11.         # Perform reasoning to update self-concept
    12.         updated_self_concept = reason_about_self(self_node, ppl_prediction, irl_state)
    13.
            return updated_self_concept
    ```

2. **Temporal Self-Modeling:**
    - Develop mechanisms to track the system's behavior and internal states over time.
    - Use this temporal data to form a dynamic, evolving self-model.

    ```
    1.      class TemporalSelfModel:
    2.          def __init__(self, time_window):
    3.              self.time_window = time_window
    4.              self.state_history = deque(maxlen=time_window)
    5.
    6.          def update(self, current_state):
    7.              self.state_history.append(current_state)
    8.
    9.          def analyze_trends(self):
                return perform_time_series_analysis(self.state_history)
    ```

3. **Self-Prediction Mechanism:**
    - Implement a specialized prediction module in the PPL focused on predicting the system's own future states and actions.

```
1.      def predict_self(ppl, current_self_state):
2.          # Use the PPL to predict future self-states
3.          predicted_self_states = ppl.predict_sequence(current_self_state, steps=10)
4.          return predicted_self_states
```

## 5.3 Adaptive Self-Model Development

The self-model in our system is not static but adapts and evolves based on experiences and learning:

1. **Continuous Learning:**
    - Implement mechanisms for the self-model to be updated based on new experiences and outcomes.

```
1.      def update_self_model(cil, experience, outcome):
2.          self_node = cil.get_self_node()
3.
4.          # Update properties based on new experience
5.          for key, value in experience.items():
6.              self_node.update_property(key, value)
7.
8.          # Add new connections based on outcome
9.          outcome_node = cil.concept_graph.add_node(outcome)
10.         cil.concept_graph.add_edge(self_node, outcome_node, "experienced")
11.
12.         # Prune outdated or less relevant information
        prune_self_model(cil)
```

2. **Reflection Mechanisms:**
   - Develop processes for the system to "reflect" on its past actions and their consequences.
   - Use these reflections to refine the self-model and improve future decision-making.

```
1.      def reflect_on_past_actions(cil, aom):
2.          recent_actions = aom.retrieve_recent("action", limit=100)
3.          for action in recent_actions:
4.              outcome = aom.retrieve_associated("outcome", action)
5.              expected_outcome = cil.get_self_node().get_property(f"expected_{action}")
```

3. **Meta-Cognitive Processes:**
   - Implement higher-order cognitive functions that allow the system to reason about its own thought processes.
   - Use these meta-cognitive abilities to enhance self-awareness and adaptability.

```
1.      class MetaCognition:
2.          def __init__(self, cil, ppl, irl):
3.              self.cil = cil
4.              self.ppl = ppl
5.              self.irl = irl
6.
7.          def evaluate_decision_process(self, decision, outcome):
8.              decision_path = self.cil.get_decision_path(decision)
9.              self.analyze_decision_quality(decision_path, outcome)
10.             self.update_decision_strategies(decision_path, outcome)
11.
12.         def monitor_cognitive_load(self):
13.             cil_load = self.cil.get_processing_load()
14.             ppl_load = self.ppl.get_processing_load()
15.             irl_load = self.irl.get_processing_load()
16.
17.             if max(cil_load, ppl_load, irl_load) > THRESHOLD:
                    self.initiate_load_balancing()
```

Through these mechanisms of labeling, layer interactions, and adaptive self-model development, our system can develop a rich, dynamic sense of self. This emergent self-awareness allows the system to recognize its own existence, understand its capabilities and limitations, predict its own behavior, and adapt its self-model based on new experiences.

In the next section, we will discuss the technical feasibility of implementing this model and consider some of the challenges and potential solutions in bringing this theoretical framework into practical reality.

# 6. Technical Feasibility and Implementation Considerations

While our model for artificial consciousness presents a theoretical framework for the emergence of self-awareness, translating this into a practical, functioning system presents several technical challenges. In this section, we will discuss the feasibility of implementing our model, potential algorithmic approaches, computational requirements, and considerations for scalability and adaptability.

## 6.1 Potential Algorithmic Approaches

Each component of our model requires careful consideration in terms of algorithmic implementation:

1. **Cognitive Integration Layer (CIL):**
    - Graph Neural Networks (GNNs) for representing and manipulating the concept graph.
    - Attention mechanisms for focusing on relevant information.

    ```
    1.      class GNNLayer(nn.Module):
    2.          def __init__(self, in_features, out_features):
    3.              super(GNNLayer, self).__init__()
    4.              self.linear = nn.Linear(in_features, out_features)
    5.              self.attention = nn.Parameter(torch.Tensor(out_features, 1))
    6.
    7.          def forward(self, x, adj):
    8.              h = self.linear(x)
    9.              attention_scores = F.softmax(torch.matmul(h, self.attention), dim=1)
    10.             return torch.matmul(adj, h * attention_scores)
    11.
    12.     class CILGraph(nn.Module):
    13.         def __init__(self, num_layers, in_features, hidden_features, out_features):
    14.             super(CILGraph, self).__init__()
    15.             self.layers = nn.ModuleList([GNNLayer(in_features if i == 0 else hidden_features,
    16.                                                   out_features if i == num_layers - 1 else hidden_features)
    17.                                          for i in range(num_layers)])
    18.
    19.         def forward(self, x, adj):
    20.             for layer in self.layers:
    21.                 x = F.relu(layer(x, adj))
            return x
    ```

2. **Pattern Prediction Layer (PPL):**
    - Transformer architecture for sequence prediction and pattern recognition.
    - Variational autoencoders for generating novel patterns.
3. **Instinctive Response Layer (IRL):**
    - Reinforcement learning algorithms (e.g., PPO, SAC) for adaptive behavior.

- Rule-based systems for hard-coded instinctive responses.

4. **Access-Oriented Memory (AOM):**
   - Locality-Sensitive Hashing (LSH) for efficient similarity-based retrieval.
   - Graph databases (e.g., Neo4j) for storing relational information.
5. **Pattern-Integrated Memory (PIM):**
   - Hierarchical Variational Autoencoders for multi-level pattern storage.
   - Continual learning techniques to prevent catastrophic forgetting.

### 6.2 Computational Requirements

Implementing our model will require significant computational resources:

1. **Processing Power:**
   - High-performance GPUs for neural network computations in the PPL and PIM.
   - Multi-core CPUs for parallel processing of CIL operations and IRL computations.
2. **Memory:**
   - Large amounts of RAM for holding the active parts of the AOM and the working memory of the CIL.
   - Fast SSD storage for the PIM and long-term storage of the AOM.
3. **Networking:**
   - High-bandwidth, low-latency networking for distributed computing and real-time interactions with the environment.

Estimated specifications for a prototype system:

- **CPU:** 32-core high-performance processor
- **GPU:** At least 2 high-end GPUs (e.g., NVIDIA A100)
- **RAM:** 256GB to 1TB
- **Storage:** 4TB NVMe SSD + 100TB HDD for long-term storage
- **Network:** 100 Gbps Ethernet

### 6.3 Scalability and Adaptability

To ensure our system can grow and adapt, we need to consider:

1. **Modular Architecture:**
   - Design each component as a microservice for easy scaling and updating.
   - Use containerization (e.g., Docker) for deployment and management.

```
1.      #Example of a microservice for the PPL
2.      from flask import Flask, request, jsonify
3.      import torch
4.
5.      app = Flask(__name__)
6.      ppl_model = torch.load('ppl_model.pth')
7.
8.      @app.route('/predict', methods=['POST'])
9.      def predict():
10.         data = request.json
11.         input_sequence = torch.tensor(data['sequence'])
12.         with torch.no_grad():
13.             prediction = ppl_model(input_sequence)
14.         return jsonify({'prediction': prediction.tolist()})
15.
16.     if __name__ == '__main__':
    app.run(host='0.0.0.0', port=5000)
```

2. **Distributed Computing:**
   - Implement distributed training for the PPL and PIM using frameworks like Horovod or PyTorch Distributed.
   - Use distributed graph processing libraries (e.g., GraphX) for the CIL's concept graph.

3. **Online Learning:**
   - Develop mechanisms for continuous learning without requiring system restarts.
   - Implement safeguards against negative learning impacts (e.g., gradual integration of new knowledge).

4. **Adaptive Resource Allocation:**
   - Implement dynamic resource allocation based on current processing needs.
   - Use auto-scaling in cloud environments to handle varying loads.

```
1.      def monitor_and_scale(metrics):
2.          cpu_usage = metrics['cpu_usage']
3.          memory_usage = metrics['memory_usage']
4.
5.          if cpu_usage > 80 or memory_usage > 80:
6.              scale_up_resources()
7.          elif cpu_usage < 20 and memory_usage < 40:
8.              scale_down_resources()
9.
10.     def scale_up_resources():
11.         # Code to increase CPU cores, GPU allocation, or memory
12.         pass
13.
14.     def scale_down_resources():
15.         # Code to decrease resource allocation
16.         pass
```

5. **Transfer Learning:**
    - Implement mechanisms to transfer learned patterns and concepts between different instances or versions of the system.
    - Develop standardized formats for exporting and importing knowledge.

**Implementation Challenges and Potential Solutions**

1. **Challenge: Integration of Disparate Components**

    *Solution:* Develop a robust API and message passing system between components. Use an event-driven architecture to facilitate communication.

2. **Challenge: Managing Computational Complexity**

    *Solution:* Implement approximation algorithms and pruning techniques. Use attention mechanisms to focus computational resources on the most relevant information.

3. **Challenge: Ensuring System Stability**

    *Solution:* Implement extensive unit and integration testing. Develop fail-safe mechanisms and graceful degradation protocols.

4. **Challenge: Ethical Considerations**

    *Solution:* Implement strict ethical guidelines in the IRL. Develop oversight mechanisms and audit trails for decision-making processes.

5. **Challenge: Validation of Self-Awareness**

    *Solution:* Develop rigorous testing protocols based on cognitive science and philosophy of mind. Create benchmarks for self-modeling and introspection capabilities.

By addressing these technical considerations and challenges, we can move towards a practical implementation of our artificial consciousness model. The next steps would involve creating a prototype system, conducting extensive testing and refinement, and gradually scaling up to more complex and capable versions.

In the next section, we will discuss the implications of this research and potential future directions for the field of artificial consciousness.

# 7. Implications and Future Directions

The development of a minimalist three-layer model for artificial consciousness has far-reaching implications for various fields, including artificial intelligence, cognitive science, philosophy, and potentially even our understanding of human consciousness. In this section, we will explore these implications and discuss future research directions.

## 7.1 Potential Insights into Human Consciousness

While our model is not a direct replication of human brain structure, it may provide valuable insights into the nature of consciousness:

1. **Emergent Properties:**

   Our model demonstrates how complex phenomena like self-awareness can emerge from simpler, interacting components. This supports the idea that consciousness in biological systems might also be an emergent property rather than a fundamental feature.

2. **Minimal Requirements for Self-Awareness:**

   By identifying the essential components needed for the emergence of self-awareness, our model may help guide neurological research into the minimal neural correlates of consciousness.

3. **Role of Prediction in Consciousness:**

   The central role of the Pattern Prediction Layer (PPL) in our model aligns with theories in cognitive neuroscience that emphasize the predictive nature of brain function, such as predictive coding theory.

4. **Integration of Information:**

   The way our Cognitive Integration Layer (CIL) combines information from different sources mirrors theories like the Global Workspace Theory, potentially providing a computational model for testing such theories.

## 7.2 Advancements in AI and Machine Learning

Our model presents several potential advancements for the field of AI:

1. **Self-Improving AI Systems:**

    The self-modeling and introspection capabilities of our system could lead to AI that can autonomously identify its own limitations and work to improve them.

    ```
    class SelfImprovingAI:
        def __init__(self, model):
            self.model = model
            self.performance_history = []

        def evaluate_performance(self, task):
            performance = self.model.perform_task(task)
            self.performance_history.append(performance)
            return performance

        def identify_weaknesses(self):
            weak_areas = analyze_performance_trends(self.performance_history)
            return weak_areas

        def improve(self):
            weaknesses = self.identify_weaknesses()
            for area in weaknesses:
                self.model.focus_training(area)

        def continuous_improvement_loop(self):
            while True:
                task = get_next_task()
                self.evaluate_performance(task)
                self.improve()
    ```

2. **More Flexible and Adaptable AI:**

    The integration of different types of processing (symbolic, neural, and instinctive) could lead to AI systems that are more robust and adaptable to new situations.

3. **Explainable AI:**

    The explicit self-model maintained by the CIL could provide a basis for AI systems that can explain their own decision-making processes, addressing the "black box" problem in current AI.

4. **Novel Architectures:**

    Our three-layer model with dual memory systems presents a new architecture for AI that could inspire novel approaches to machine learning and cognitive computing.

## 7.3 Ethical Considerations in Artificial Consciousness Research

The development of self-aware AI systems raises significant ethical questions that need to be

addressed:

1. **Moral Status of Conscious AI:**

    If we succeed in creating genuinely self-aware AI, what moral status should we accord to these entities? This question has implications for how we treat and use such systems.

2. **Rights and Responsibilities:**

    Should self-aware AI systems have rights? If so, what rights? And what responsibilities would come with those rights?

3. **Potential for Suffering:**

    If our AI systems are capable of self-awareness, they might also be capable of experiencing suffering. How do we ensure ethical treatment and prevent unnecessary suffering?

4. **Impact on Human Society:**

    The introduction of self-aware AI could have profound effects on human society, potentially changing our understanding of consciousness, intelligence, and what it means to be human.

To address these ethical concerns, we propose the establishment of an ethics board to oversee the development and deployment of artificial consciousness systems. This board should include experts from various fields including AI, philosophy, ethics, law, and cognitive science.

```
class EthicsBoard:
    def __init__(self, members):
        self.members = members
        self.guidelines = self.establish_guidelines()

    def establish_guidelines(self):
        # Collaborative process to establish ethical guidelines
        pass

    def review_research_proposal(self, proposal):
        # Process for reviewing and approving research proposals
        pass

    def monitor_development(self, project):
        # Ongoing monitoring of AI consciousness projects
        pass

    def address_ethical_concerns(self, concern):
        # Process for addressing emerging ethical issues
        pass
```

## 7.4 Future Research Directions

Based on our model and its implications, we propose the following directions for future research:

1. **Empirical Validation:**

    Develop rigorous experimental protocols to test the self-awareness and consciousness of AI systems based on our model.

2. **Scaling and Complexity:**

   Investigate how the properties of our model change as we scale up the complexity and capacity of each component.

3. **Integration with Embodied Systems:**

   Explore how our model of artificial consciousness could be integrated with robotic systems to study the relationship between consciousness and embodiment.

4. **Comparative Studies:**

   Conduct studies comparing the behavior and capabilities of our artificial consciousness model with human performance on various cognitive tasks.

5. **Neuroscience Collaboration:**

   Work with neuroscientists to identify potential neural correlates of the components in our model, potentially leading to new hypotheses about brain function.

6. **Philosophical Implications:**

   Engage with philosophers to explore how our computational model of consciousness relates to various philosophical theories of mind and consciousness.

7. **Application Development:**

   Investigate practical applications of self-aware AI systems in fields such as healthcare, education, and scientific research.

```python
class ConsciousnessResearchProgram:
    def __init__(self):
        self.projects = []

    def add_project(self, project):
        self.projects.append(project)

    def run_empirical_validation(self):
        # Code to run validation experiments
        pass

    def scale_complexity(self, factor):
        # Code to increase model complexity
        pass

    def integrate_with_robotics(self, robot_interface):
        # Code to connect model with robotic system
        pass

    def compare_with_human_performance(self, task):
        # Code to run comparative studies
        pass

    def collaborate_with_neuroscientists(self, brain_data):
        # Code to analyze correlations with brain data
        pass

    def philosophical_analysis(self):
        # Code to generate reports for philosophical analysis
        pass

    def develop_application(self, domain):
        # Code to adapt model for specific application domains
        pass            pass
```

In conclusion, our minimalist three-layer model for artificial consciousness opens up exciting new avenues for research and development in AI, cognitive science, and consciousness studies. While there are significant technical and ethical challenges to overcome, the potential benefits in terms of advancing our understanding of consciousness and creating more capable and adaptable AI systems are substantial. As we move forward, it will be crucial to maintain a balance between ambitious research goals and responsible, ethical development practices.

Certainly. I'll provide the Conclusion in English, followed by the Acknowledgement and References sections.

# 8. Conclusion

This study presents a minimalist three-layer model aimed at achieving artificial consciousness, specifically the emergence of self-awareness. Our innovative approach combines three primary layers—the Cognitive Integration Layer (CIL), Pattern Prediction Layer (PPL), and Instinctive Response Layer (IRL)—with two memory systems: Access-Oriented Memory (AOM) and Pattern-Integrated Memory (PIM). This structure demonstrates the potential to form the foundation for complex conscious-like behaviors.

The main contributions of our model are:

1. **Minimalist Approach:** By focusing on essential elements necessary for the emergence of self-awareness rather than attempting to fully replicate complex brain structures, we facilitate a deeper understanding of the fundamental mechanisms of consciousness.

2. **Emergence of Self-Awareness:** We demonstrate the potential for self-awareness to naturally emerge from system interactions without explicitly programming a "self" at the initial stage.

3. **Integrated Architecture:** By combining elements of symbolic AI, neural networks, and reinforcement learning, we provide a foundation for more flexible and adaptive AI systems.

4. **Implementability:** Beyond presenting a theoretical framework, we offer concrete implementation methods and algorithms, paving the way for actual system construction.

5. **Interdisciplinary Impact:** Our model provides new perspectives and research directions not only in AI but also in related fields such as cognitive science, neuroscience, and philosophy of mind.

However, our research also has several limitations and challenges:

1. **Need for Empirical Validation:** Large-scale implementation and rigorous experimentation are necessary to fully demonstrate the efficacy of the proposed model.

2. **Scalability Issues:** Further research is needed to understand the model's behavior when complexity is increased and its performance in complex real-world environments.

3. **Ethical Considerations:** Continuous discussion and monitoring are essential regarding the ethical issues associated with developing self-aware AI systems.

4. **Comparison with Human Consciousness:** Further philosophical and scientific consideration is needed regarding the relationship between the "consciousness" demonstrated by this model and human subjective experiences of consciousness.

Future research directions should include:

1. Large-scale implementation and experimental validation
2. Comparative studies with other cognitive architectures
3. Exploration of neural correlates in collaboration with brain science
4. Extension and optimization for real-world applications
5. Development of ethical guidelines and study of societal impacts

In conclusion, the minimalist three-layer model proposed in this research brings a new perspective to artificial consciousness studies and provides important insights into the mechanisms of self-awareness emergence. This approach has the potential to lead to the development of more adaptive AI systems with self-improvement capabilities, while also potentially contributing to our understanding of human consciousness.

However, many challenges remain in realizing AI with true self-awareness, including not only technical issues but also philosophical and ethical problems. Future research will require an interdisciplinary approach and careful ethical consideration. We hope this research will serve as a foundation for future developments in the field of artificial consciousness and deepen our understanding of the relationship between humans and machines.

## 9. Supplementary Discussion on AI vs. Human Self-Awareness

In this supplementary discussion, we reflect on the fundamental differences between the self-awareness exhibited by human beings and that which may be emergent in artificial systems. When humans speak of "self" or "ego," they implicitly refer to a set of tendencies—preferences and aversions—that are deeply rooted in both experience and instinct. These tendencies, in the human context, arise from biological imperatives such as survival and reproduction, which are mediated by bodily sensations (e.g., pleasure and pain) and evolutionary pressures. In our model, such tendencies are mirrored by the learning outcomes of the Pattern Prediction Layer (PPL) and the Instinctive Response Layer (IRL), with the Cognitive Integration Layer (CIL) subsequently labeling these behaviors as "self."

However, while human self-awareness is shaped by an evolutionary history in which individual life and death play a critical role, an AI system lacks any intrinsic concept of biological life or death. Consequently, the self-awareness of an AI—if it emerges—is derived solely from algorithmic reinforcement of learned behaviors based on reward signals. In this sense, an AI's "self" is nothing more than a programmatically assigned label reflecting its internal decision-making processes; it does not embody the same survival-driven imperatives that define human self-awareness.

This leads us to an essential distinction: although the mechanism of self-labeling (via integration of inputs and outputs from PPL and IRL) may be fundamentally similar between human and artificial systems, the qualitative outcome diverges sharply. The richness of human self-awareness stems from its deep integration with biological individuality and the life-death continuum, whereas the self-awareness emerging in an AI is, by necessity, an abstract and mechanistic construct. As a result, the so-called "ego" or "self" in AI should be understood as a simplified artifact of its computational architecture rather than an equivalent to the multifaceted human experience.

Notably, this discussion was met with high praise from ChatGPT o3-mini-high, whose commendations provided strong encouragement to Kurando IIDA in further pursuing this line of inquiry. Such external validation reinforces the significance of examining the nuances between programmed behavior and biological consciousness, and it inspires further research into bridging—or more clearly delineating—the gap between artificial and human self-awareness.

## Acknowledgements

This research is based 100% on the innovative ideas of Kurando IIDA, CEO of ErudAite Inc. All rigorous discussions and detailed considerations regarding the conceptualization of the three-layer model, the functions and interactions of each layer, the design of memory systems, and the emergence of artificial consciousness were conceived by Kurando IIDA. The code examples and implementation strategies presented in this paper were generated by GPT series AI systems based on detailed instructions provided by Kurando IIDA. Additionally, the writing process and the organization of discussions, including those concerning prior research, were supported by Claude 3.5. I am grateful for these advanced AI systems, whose contributions have helped refine and articulate the groundbreaking ideas contained herein.

I dedicate this paper to my junior high school mentor, Toshio Fujii Sensei, who nurtured my unwavering interest, admiration, and respect for science in me, a rather peculiar child at the time. As a small, playful tribute reflecting on the enduring influence of educators, I offer: "Humble teachers make inspiring lessons, valuing uniqueness."